\documentclass[twocolumn,showpacs,aps,prl,superscriptaddress,,amssymb,floatfix,preprintnumbers]{revtex4}

\usepackage{graphicx}
\usepackage{dcolumn}
\usepackage{amsmath}
\usepackage{epsfig}
\usepackage{color}
\input babarsym

\newcommand{\BABARPubYear}    {07}
\newcommand{\BABARPubNumber}  {053}

\newcommand{\SLACPubNumber} {12938}

\def\figurebox#1#2#3{%
    \def\arg{#3}%
    \ifx\arg\empty
    {\hfill\vbox{\hsize#2\hrule\hbox to #2{\vrule\hfill\vbox to #1{\hsize#2\vfill}\vrule}\hrule}\hfill}%
    \else
    {\hfill\epsfbox{#3}\hfill}%
    \fi}

\begin{document}

\preprint{\babar-PUB-\BABARPubYear/\BABARPubNumber} 
\preprint{SLAC-PUB-\SLACPubNumber} 


\title{
{\large \bf
A study of $\mathbf{\overline{B} \to \Xi_c \bar{\Lambda}_c^-}$ and $\mathbf{\overline{B} \to \Lambda_c^+ \bar{\Lambda}_c^- \Kb}$ decays at \babar}
}

%
\author{B.~Aubert}
\author{M.~Bona}
\author{D.~Boutigny}
\author{Y.~Karyotakis}
\author{J.~P.~Lees}
\author{V.~Poireau}
\author{X.~Prudent}
\author{V.~Tisserand}
\author{A.~Zghiche}
\affiliation{Laboratoire de Physique des Particules, IN2P3/CNRS et Universit\'e de Savoie, F-74941 Annecy-Le-Vieux, France }
\author{J.~Garra~Tico}
\author{E.~Grauges}
\affiliation{Universitat de Barcelona, Facultat de Fisica, Departament ECM, E-08028 Barcelona, Spain }
% ...
\collaboration{The \babar\ Collaboration}
\noaffiliation

\date{\today}

\begin{abstract}
We report measurements of $B$-meson decays into two- and three-body final states
containing two charmed baryons using a sample of 230 million \upsbb decays.
We find significant signals in two modes, measuring branching fractions
  $\BR(\BmtoLcLcKm) = (1.14 \pm 0.15 \pm 0.17 \pm 0.60)\times10^{-3}$ and 
  $\BR(\BmtoXiczLc)\times \BR(\XcztoXipi) = (2.08 \pm 0.65 \pm 0.29 \pm 0.54)\times10^{-5}$, 
where the uncertainties are statistical, systematic, and from the branching fraction
$\BR(\Lctopkpi)$, respectively. We also set upper limits at the 90\% confidence level on two other modes:
  $\BR(\Bzb\to\Xi_c^+\aLc)\times\BR(\Xi_c^+\to\Xi^-\pi^+\pi^+) < 5.6\times 10^{-5}$ and
  $\BR(\Bzb\to\Lc\aLc\Kzb) < 1.5 \times 10^{-3}$.
We observe structure centered at an invariant mass of 2.93~GeV$/c^2$
in the $\Lc K^-$ mass distribution of the decay \BmtoLcLcKm.
\end{abstract}


\maketitle

Bottom $(B)$ mesons are heavy enough to decay into charmed baryons, and do so at a rate of roughly 5\%~\cite{ref:PDGbook,ref:babarLc}. The dominant decay mechanism is via $\b \to \c \Wm$ transitions, with \Wm\ coupling to $\cbar\s$ or $\ubar\d$~\cite{charge_conjugation}, both of which are Cabibbo-allowed. Theoretical predictions for the branching fractions of \B\ mesons to baryon-antibaryon pairs have been made within the diquark model~\cite{ref:Chernyak:1990ag} and with QCD sum rules~\cite{ref:Ball:1990fw}. These suggest that decays to two charmed baryons ($\Bb \to X_{c1} \bar{X}_{c2}$) and to one charmed baryon and one light baryon ($\Bb \to X_{c1} \bar{X}_{2}$) have comparable branching fractions, of the order of $10^{-3}$ for individual modes. 

 Several inclusive measurements of $B$-meson decays to charmed baryons have been made~\cite{ref:PDGbook}. In particular, the \babar\ Collaboration recently performed an inclusive analysis of $\Lambda_c^+$ production in which flavor tag information was used to identify whether the $\Lambda_c^+$ came from a \B\ or a \Bb\ meson~\cite{ref:InclLc}. It was found that about a third of all \Lc were from \B\ mesons with anti-correlated flavor content (i.e. $\overline{b} \to c$ rather than $b \to c$ transitions), consistent with a substantial rate of $\b\to \c \cbar\s$ decays. Inclusive studies of the $\Xi_c^0$ and $\Lambda_c^+$ momentum spectrum~\cite{ref:Aubert:2005cu,ref:belleLc,ref:babarLc} also support a substantial rate of baryonic $b\to\ccbar s$ decays such as $\Bm\to\Xi_c^0\aLc$. However, inclusive studies alone cannot fully establish this, since the momentum distributions can also be reproduced with carefully tuned sums of $b \to c \bar{u} d$ processes. Therefore, exclusive measurements are needed. These require very large samples of $B$-meson decays and have only recently become feasible.

 The Belle Collaboration has reported results on $B$ decays to final states with two charmed baryons in both two- and three-body modes~\cite{ref:twobodycharm,ref:threebodycharm}. 
They measured 
  $\BR(\Bm\to\Xcz\aLc) \times \BR(\Xcz\to\Xim\pip) = (4.8^{+1.0}_{-0.9} \pm 1.1 \pm 1.2) \times 10^{-5}$ and
  $\BR(\Bzb\to\Xcp\aLc) \times \BR(\Xcp\to\Xim\pip\pip) = (9.3^{+3.7}_{-2.8} \pm 1.9 \pm 2.4) \times 10^{-5}$~\cite{ref:twobodycharm}.
Assuming that $\BR(\Xcz\to\Xim\pip)$ and $\BR(\Xcp\to\Xim\pip\pip)$ are of the order of 1\%--2\%~\cite{ref:k.k.sharma}, these results are compatible with the prediction that $\BR(\Bm\to\Xcz\aLc)$ and $\BR(\Bzb\to\Xcp\aLc)$ are $\mathcal{O}(10^{-3})$. This is in stark contrast to the branching fractions of singly charmed decays, such as that of $\Bzb \to \Lc \bar{p}$ which is $(2.2 \pm 0.8)\times10^{-5}$, smaller by two orders of magnitude~\cite{ref:twobodysinglecharm}. The branching fractions of the three-body processes $\Bb\to\Lc\aLc\Kb$ were also found to be large:
$\BR(\Bm\to\Lc\aLc\Km) = (0.65^{+0.10}_{-0.09}\pm0.11\pm0.34)\times10^{-3}$
$\BR(\Bzb\to\Lc\aLc\Kzb) = (0.79^{+0.29}_{-0.23} \pm 0.12 \pm 0.42) \times 10^{-3}$~\cite{ref:threebodycharm}. Explanations for these widely varying values have been proposed~\cite{ref:Cheng2005vd, ref:Chen:2006fs}. It was suggested that a kinematic suppression may apply to decays in which the two baryons have high relative momentum, since this requires the exchange of two high-momentum gluons. The rate of $\Bb\to\Lc\aLc\Kb$ decays could also be enhanced by final-state interactions, or by intermediate charmonium resonances. 

 In this paper, we present measurements of the branching fraction of the decays
$\BmtoLcLcKm$,
$\BmtoXiczLc$, 
$\BztoXicpLc$, 
and $\Bzb\to\Lc\aLc\Kzb$, and investigate three-body decays for the possible presence of intermediate resonances. The data were collected with the \babar\ detector~\cite{ref:Detector} at the \pep2\ asymmetric-energy \epem\ storage rings and represent an integrated luminosity of approximately 210 \invfb\ collected at a center-of-mass energy $\sqrt{s} = 10.58$~\gev, corresponding to the mass of the \FourS\ resonance. 
The \babar\ detector is a magnetic spectrometer with 92\% solid angle tracking coverage
in the center-of-mass frame.
Charged particles are detected and their momenta are measured in a
five-layer double-sided silicon vertex tracker and a forty-layer drift chamber, both
operating in a 1.5~T magnetic field. Charged particle identification (PID) is
provided by the average energy loss (d$E$/d$x$) in the tracking devices and by an internally
reflecting ring-imaging Cherenkov detector. Photons are detected with a CsI(Tl)
electromagnetic calorimeter. The instrumented flux return for the solenoidal magnet provides
muon identification. 
 Simulated events with $B$ mesons decaying into the relevant final states are generated with \evtgen~\cite{bib:evtgen} and {\tt PYTHIA}~\cite{bib:pythia}, while {\tt GEANT4}~\cite{bib:geant4} is used to simulate the detector response. 
Inclusive Monte Carlo (MC) samples of 
$\Upsilon(4S)$ and $\epem \to \qqbar$ $(q=u,d,s,c)$ events at $\sqrt{s} = 10.58$~\gev
are also used, corresponding to more than 1.5~times the integrated luminosity of the data.

 The $\Lc$ candidates are reconstructed in the three decay modes $p\Km\pip$, $p\KS$, and $\Lz\pip$; \Xcz candidates in the two decay modes $\Xim\pip$ and $\Lz\Km\pip$; and \Xcp candidates in the decay mode $\Xim\pip\pip$. 
 We begin by reconstructing the long-lived strange hadrons: $\KS\to\pi^+\pi^-$ and $\Lambda\to p \pi^-$ candidates are reconstructed from two oppositely charged tracks, and $\Xim\to\Lambda\pi^-$ from a \Lz candidate and a negatively charged track. In each case, we fit the daughters to a common vertex and compute their invariant mass. The mass is required to be within $3\sigma$ of the central value, where $\sigma$ is the experimental resolution and is approximately 4.0, 4.5, and 6.0~MeV$/c^2$ for \KS, \Lz, and \Xim, respectively. Candidates with a $\chi^2$ probability below $10^{-4}$ are rejected. For \Lz candidates, we also require the daughter proton to satisfy PID criteria. The mass of the \KS, \Lz, or \Xim candidate is constrained to its nominal value~\cite{ref:PDGbook} for subsequent fits. 

 We suppress background by requiring the transverse displacement between the event and decay vertices to be greater than 0.2~centimeters for \KS, \Lz, and \Xim, each of which travels several centimeters on average. We also require that the scalar product of the displacement and momentum vectors of each hadron be greater than zero, and that the transverse component of the displacement vector of a \Xim candidate be smaller than that of its \Lz daughter.

\begin{figure}[tb]
  \begin{center}
    \epsfig{file=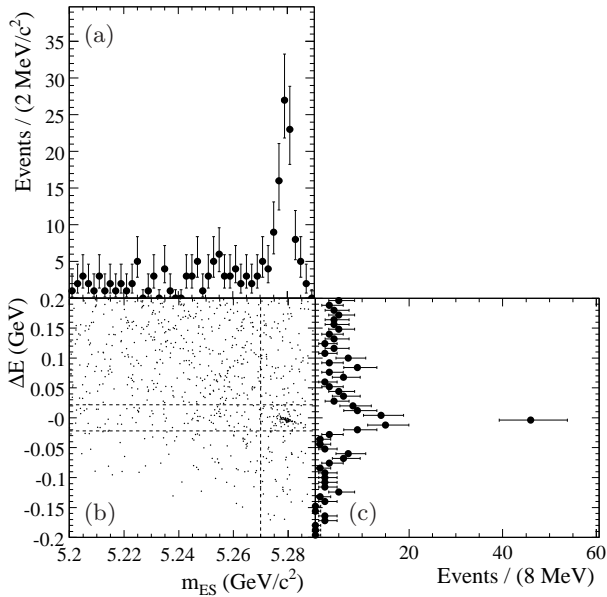, width=0.97\columnwidth}
  \begin{picture}(0,0)
    \put( -210, 213){(a)}
    \put( -210,  32){(b)}
    \put( -110,  32){(c)}
  \end{picture}
  \end{center}
  \caption{The \mes\ and \de distributions for \BmtoLcLcKm candidates, summing over five different final states. Plot~(b) shows the scatterplot of \mes\ vs.\ \de, and (a) and (c) show the \mes\ and \de projections for $|\de|<$ 0.022 \gev and for $\mes>$ 5.27 \gevcc, respectively. The dashed horizontal and vertical lines in~(b) indicate the signal regions used for the projection in~(a) and~(c), respectively.
  }
  \label{data_plot}
\end{figure}

\begin{table*}
\caption{Fitted signal yield, detection efficiency $\varepsilon$, significance $S$, measured branching fraction $\cal{B}$, and (for $S<2$) the upper limit on $\cal{B}$ for each decay mode. The uncertainties on $\cal{B}$ are statistical, systematic, and the uncertainty from the branching fraction $\BR(\Lctopkpi)$. For final states containing $\Xi_c^0$ or $\Xi_c^+$, $\BR$ includes a factor of $\BR(\Xi_c^0\to\Xim\pip)$ or $\BR(\Xi_c^+\to\Xim\pip\pip)$, respectively.}
\label{tab:Brresult}
\begin{center}
\begin{tabular}{l c c c c c}
\hline \hline
\multicolumn{1}{l}{Decay Mode}&Signal Yield &$\varepsilon$$(\%)$ & $S$ & $\cal{B}$ & Upper Limit on $\cal{B}$ \\ 
\hline \noalign{\vskip1pt}
$\BmtoLcLcKm$                      &$74.6 \pm 9.8$ & --- & 9.6 & $(1.14\pm0.15\pm0.17\pm0.60)\times 10^{-3}$ & \\ \noalign{\vskip1pt}
\quad\Lctopkpi, \aLctopkpi         &$42.7 \pm 7.7$ & 7.1 & 7.1 & $(1.07\pm0.19\pm0.16\pm0.56)\times 10^{-3}$ & \\ \noalign{\vskip1pt}
\quad\Lctopkpi, \aLctopks          &$14.5 \pm 4.0$ & 8.8 & 5.9 & $(1.81\pm0.50\pm0.30\pm0.94)\times 10^{-3}$ & \\ \noalign{\vskip1pt}
\quad\Lctopks, \aLctopkpi          &$11.4 \pm 3.7$ & 8.8 & 4.8 & $(1.42\pm0.45\pm0.24\pm0.74)\times 10^{-3}$ & \\ \noalign{\vskip1pt}
\quad\Lctopkpi, \aLctoLzpi         &$ 2.5 \pm 1.8$ & 6.3 & 2.0 & $(0.55\pm0.40\pm0.09\pm0.28)\times 10^{-3}$ & \\ \noalign{\vskip1pt}
\quad\LctoLzpi, \aLctopkpi         &$ 3.5 \pm 2.0$ & 6.4 & 2.7 & $(0.74\pm0.43\pm0.12\pm0.38)\times 10^{-3}$ & \\
\hline \noalign{\vskip1pt}
\BmtoXiczLc                        &$14.0 \pm 4.4$ & --- & 6.4 & $(2.08\pm0.65\pm0.29\pm0.54)\times 10^{-5}$ & \\ \noalign{\vskip1pt}
\quad\XcztoXipi, \aLctopkpi        &$ 8.0 \pm 2.8$ & 4.3 & 6.1 & $(2.51\pm0.89\pm0.29\pm0.65)\times 10^{-5}$ & \\ \noalign{\vskip1pt}
\quad\XcztoLzkpi, \aLctopkpi       &$ 6.0 \pm 3.4$ & 4.5 & 2.1 & $(1.70\pm0.93\pm0.30\pm0.44)\times 10^{-5}$ & \\
\hline  \noalign{\vskip1pt}
\BztoXicpLc                        &$ 2.8 \pm 2.0$ & 2.6 & 1.8 & $(1.50\pm 1.07\pm0.20\pm0.39)\times10^{-5}$ & ~$< 5.6 \times 10^{-5}$ @ 90\% C.L. \\
\hline \noalign{\vskip1pt}
$\Bzb\to\Lc\aLc\Kzb$               &$ 3.3 \pm 2.7$ & 4.4 & 1.4 & $(0.38\pm0.31\pm0.05\pm0.20)\times10^{-3}$ & ~$< 1.5 \times 10^{-3}$ @ 90\% C.L. \\
\hline \hline
\end{tabular}
\end{center}
\end{table*}

 Next, we reconstruct the charmed baryons \Lc, \Xcz, and \Xcp in the decay modes listed previously. In each case, we fit their daughters to a common vertex, require the invariant mass of the charmed baryon candidate to be within $18\mev/c^2$ (approximately three times the experimental resolution) of the nominal mass~\cite{ref:PDGbook}, reject candidates with a $\chi^2$ probability below $10^{-4}$, and then constrain the masses to their nominal values. We also require that daughter kaons and protons of the charmed baryons satisfy the PID criteria for that hypothesis.

 We reconstruct \Bb-meson candidates in the following final states: 
$\Lc \aLc K^-$,
$\Xi_c^0 \aLc$, 
$\Xi_c^+ \aLc$, 
and $\Lc \aLc \KS$, fitting the daughters to a common vertex and
requiring that the $\chi^2$ probability is at least $10^{-4}$.
We also apply the kinematic and PID requirements mentioned above
to the $\KS$ and $K^-$ daughters of the $B$ mesons.
Because the branching fraction and efficiency are higher for $\Lc \to p\Km\pip$ than for the other $\Lc$ decay modes, we use only final states in which at least one $\Lc$ or $\aLc$ decays to $p\Km\pip$ or $\antiproton\Kp\pim$. For each $B$-meson candidate, we compute the energy-substituted mass $\mes \equiv (s/4 - {\pcm}^{2})^{1/2}$ and the energy difference $\DeltaE \equiv E^{*}_{B} - \sqrt{s}/2$, where \pcm, $E^{*}_{B}$, and $\sqrt{s}$ are the momentum and energy of the $B$ meson and the \epem collision energy, respectively, all calculated in the \epem center-of-mass frame. For a correctly reconstructed signal decay, the \mes\ distribution peaks near the nominal mass of the $B$ meson with a resolution of approximately $2.5\mev/c^2$, and \de\ peaks near zero with a resolution of $6.0$--$7.8\mev$ depending on the final state. Figure~\ref{data_plot} shows the \mes\ and \de\ distributions for \BmtoLcLcKm candidates.

 Background arises from several sources, including mis-reconstructed $B$ decays to two charmed baryons, $B$ decays to a single charmed baryon, $\epem\to\ccbar$ events containing charmed baryons, and random combinations of tracks. We use inclusive MC simulations and events from the sidebands of \mes, \de, and charmed baryon mass in data to study the background. 
We consider as background $B$-meson decays with the same final state that do not proceed via an intermediate charmed baryon---for example, $B^- \to \Xi_c^0 \bar{p} K^+ \pi^-$ misinterpreted as $B^- \to \Xi_c^0 \aLc$. Decays of this kind are distributed as signal in \mes\ and \de but have a smooth distribution for the mass spectrum of the misreconstructed charmed baryon, unlike signal decays which also peak in the charmed baryon mass. 
In studies of the $\Xi_c$ and $\Lambda_c$ mass sidebands, we find no evidence for these processes and conclude that their contribution is negligible.

 Another important source of background is feed-down from related processes.
The $B$ meson can undergo a quasi-two-body decay via an excited charmed
baryon such as $\Bb \to \Xi_c^* \aLc$, or a non-resonant multi-body decay
such as $\Bb \to \Xi_c \aLc \pi$. These events have similar distributions to
the signal for \mes\ and the charmed baryon invariant masses, but are
displaced in \de\ by an amount that depends on the final state but is
generally more than 50~MeV. We remove these backgrounds by requiring that
signal candidates satisfy $|\de|<22$~MeV. 
Finally, we require $5.2 < \mes < 5.3$~$\gev/c^2$. The average number of
reconstructed $B$ candidates per selected event varies between 1.00 and 1.14
depending on the final state. In events with more than one candidate, the
one with the smallest $|\de|$ is chosen. We verify with MC and events from
data sidebands that this does not introduce any bias in the signal
extraction. Studies of simulated events show that 1\%--3\% of signal events
are incorrectly reconstructed with one or more tracks originating from the
other $B$ in the event; this effect is taken into account implicitly by the
efficiency correction described later. 

\begin{figure}[tb]
  \begin{center}
    \epsfig{file=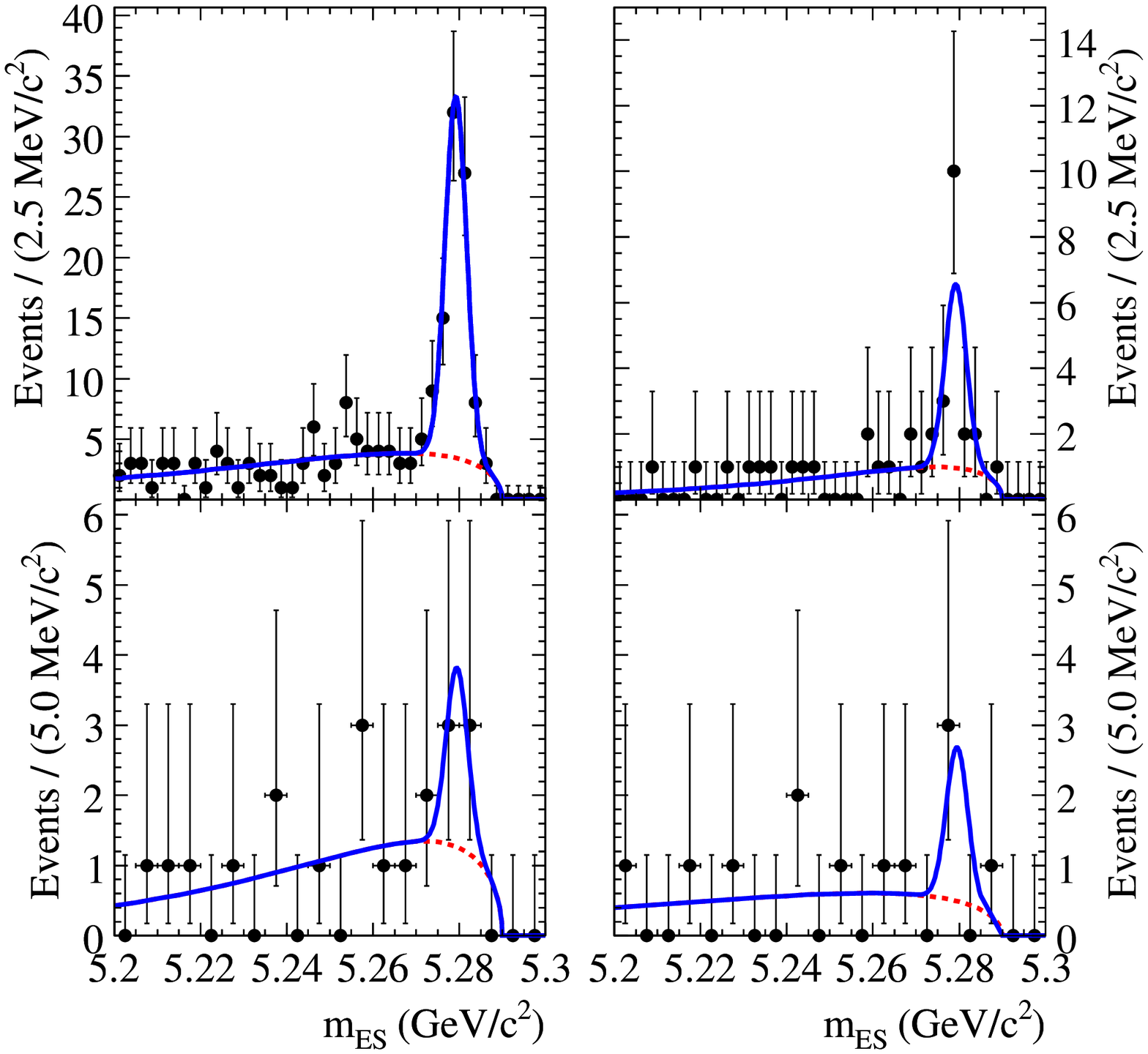, width=0.97\columnwidth}
  \begin{picture}(0,0)
    \put(-210, 200){(a)}
    \put(-100, 200){(b)}
    \put(-210,  95){(c)}
    \put(-100,  95){(d)}
  \end{picture}
  \end{center}
\caption{The fitted \mes\ distributions observed for the decay modes
  (a) $\Bm \to \Lc \aLc \Km$, combining 5 exclusive final states;
  (b) $\Bm \to \Xi_c^0 \aLc$, combining 2 exclusive final states;
  (c) \BztoXicpLc;
  (d) \BztoLcLcKz.
  Points with error bars represent the data, dashed lines the background PDF, and solid lines the sum of the signal and background PDFs.}
\label{figure:mes_Lclc_sum}
\end{figure}

The signal yields are extracted from an unbinned extended maximum likelihood fit to the \mes distribution. We use separate probability density functions (PDFs) for signal and background events. The likelihood function for the $N$ candidates in the event sample is given by
\begin{equation}
\label{eq:MLF}
{\mathcal L} = 
 \frac{e^{-(n_S+n_B)}}{N!}
 \prod_{i=1}^{N} \left( n_S \mathcal{P}_S(\mes_i) + n_B \mathcal{P}_B(\mes_i) \right)\textrm{,}
\end{equation}
where 
  $S$ here denotes the signal and $B$ the background,
  $\mathcal{P}$ is the PDF (normalized to unit integral), and
  $n$ is the yield.
The signal PDF is parameterized as a Gaussian function with $\sigma$ fixed to a value obtained from a fit to simulated signal events. The Gaussian mean is also fixed to the value obtained with simulated signal events, except for $\BmtoLcLcKm$ where there is sufficient signal in the data to fit this parameter. 
The background PDF is parameterized as an ARGUS function~\cite{ref:argus}. We allow the ARGUS shape parameter to vary within a physically reasonable range in the fit to the data.

The fitted \mes\ distributions of the four final states are shown in
Fig.~\ref{figure:mes_Lclc_sum}. Clear signals are seen in the
\BmtoLcLcKm and \BmtoXiczLc decay modes. A measure of the significance
of each peak is given by $S =
\sqrt{2\Delta\ln\cal{L}}$~\cite{ref:PDGbook}, where $\Delta\ln\cal{L}$
is the difference in likelihood (incorporating the fitting systematic
uncertainty) for fits where the signal yield is allowed to vary and where
it is fixed to zero, respectively. The results of the fits are shown in
Table \ref{tab:Brresult}. 
The efficiency is determined by applying the same analysis procedure
to simulated signal events. For the
three-body $B$-meson decays, the efficiency depends upon the distribution
in the Dalitz plane. We weight the simulated events to reproduce the
efficiency-corrected, background-subtracted distribution seen in
data for \BmtoLcLcKm. As a crosscheck, we also compute the efficiency
assuming a phase-space distribution and find a difference of less than
10\% in each case.

We then obtain each branching fraction as:
\begin{equation}
  \BR(\Bbar \to X_{c} \aLc [\Kb]) = \frac{\sum_j n_{Sj}}{{N_{\Bbar}} \sum_j \left(\varepsilon_j \prod_i{\mathcal{B}_{ij}}\right) }
  \label{eq:computeBF}
\end{equation}
where 
  $X_{c}$ is the charmed baryon ($\Lambda_c^+$, $\Xi_c^0$, or $\Xi_c^+$),
  $n_{Sj}$ is the signal yield extracted from the fit to the data for the $j^{\mathrm{th}}$ sub-mode,
  $\prod_i{\mathcal{B}_{ij}}$ is the product of the daughter branching fractions,
  $N_{\Bbar}$ is the number of neutral or charged $B$ mesons,
  and $\varepsilon_j$ is the signal detection efficiency. 
We assume equal decay rates of the \FourS\ to \BpBm\ and \BzBzb~\cite{ref:PDGbook}. 

The branching fraction $\BR(\Lc \to p \Km \pip)$ has been measured
previously to be $(5.0 \pm 1.3)\%$~\cite{ref:PDGbook}.  
Because the
branching fractions of $\Xi_c^0$ and $\Xi_c^+$ decays have not been determined experimentally,
we quote the products of the branching fractions,
$\BR(\BmtoXiczLc) \times \BR(\Xi_c^0 \to \Xi^- \pi^+)$ and
$\BR(\BztoXicpLc) \times \BR(\Xi_c^+ \to \Xi^- \pi^+ \pi^+)$.
For the $\Xi_c^0 \to \Lambda K^- \pi^+$ decay mode we
scale the measured branching fraction by the ratio
$\BR(\Xi_c^0 \to \Xi^- \pi^+)/\BR(\Xi_c^0 \to \Lambda K^- \pi^+) = 1.07 \pm 0.14$~\cite{ref:PDGbook}
so that its value can also be expressed as the product of the same two
branching fractions.

 For each decay mode, Table \ref{tab:Brresult} gives the values of 
  $n_S$,
  $\varepsilon$, 
  the significance,
  and the branching fraction.
For each mode with a significance below 2 standard deviations, we calculate the Bayesian upper limit~\cite{ref:PDGbook} on the branching fraction including systematic uncertainties and obtain
  $\BR(\Bzb\to\Lc\aLc\Kzb) < 1.5 \times 10^{-3}$ and 
  $\BR(\Bzb\to\Xi_c^+\aLc)\times\BR(\Xi_c^+\to\Xi^-\pi^+\pi^+) < 5.6\times 10^{-5}$ 
at the 90\% confidence level.

\begin{figure*}
  \begin{center}
    \begin{tabular}{ccc}
      \epsfig{file=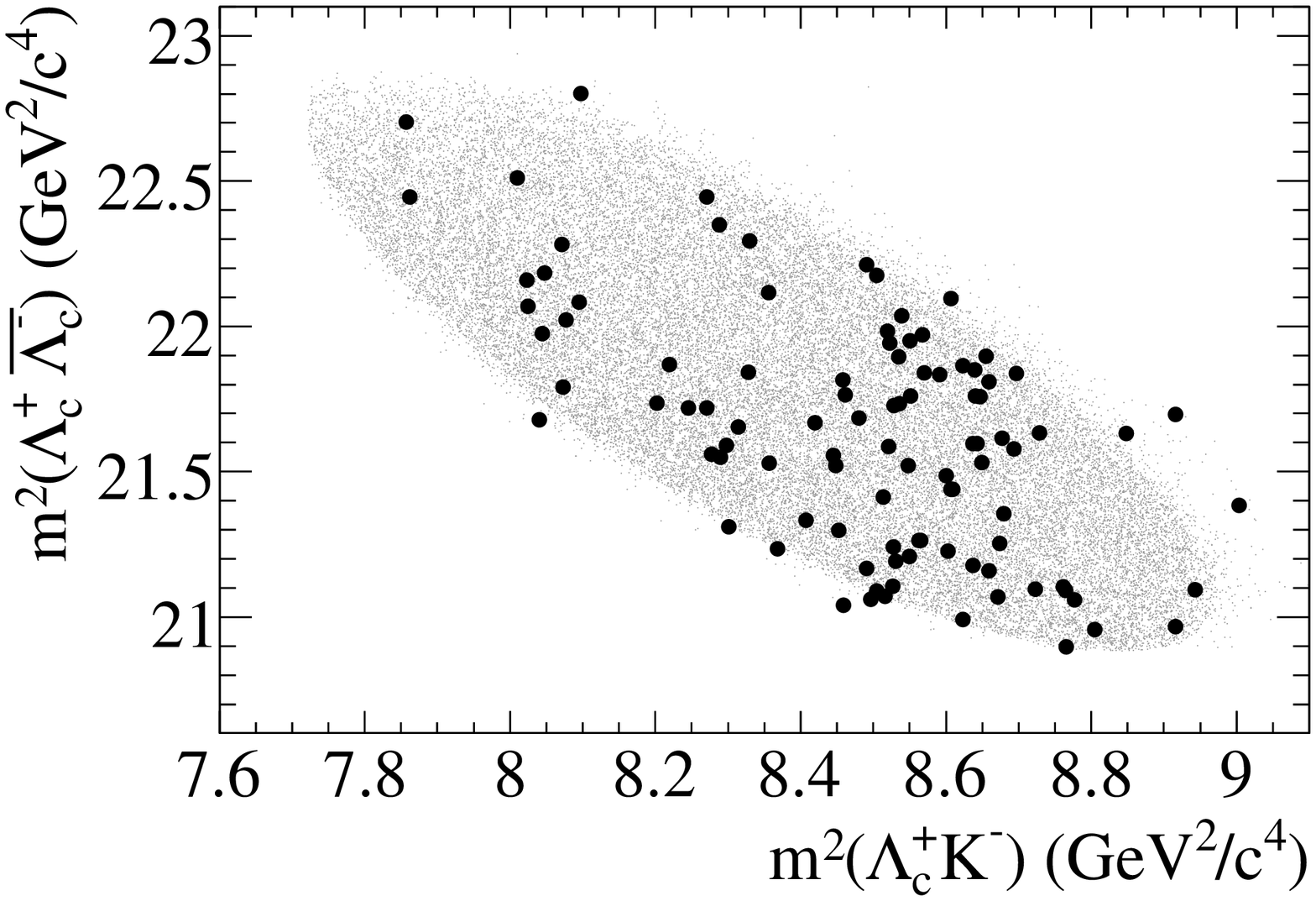, width=0.32\textwidth} 
      \begin{picture}(0,0)
	\put(-28, 95){(a)}
      \end{picture}
      &
      \epsfig{file=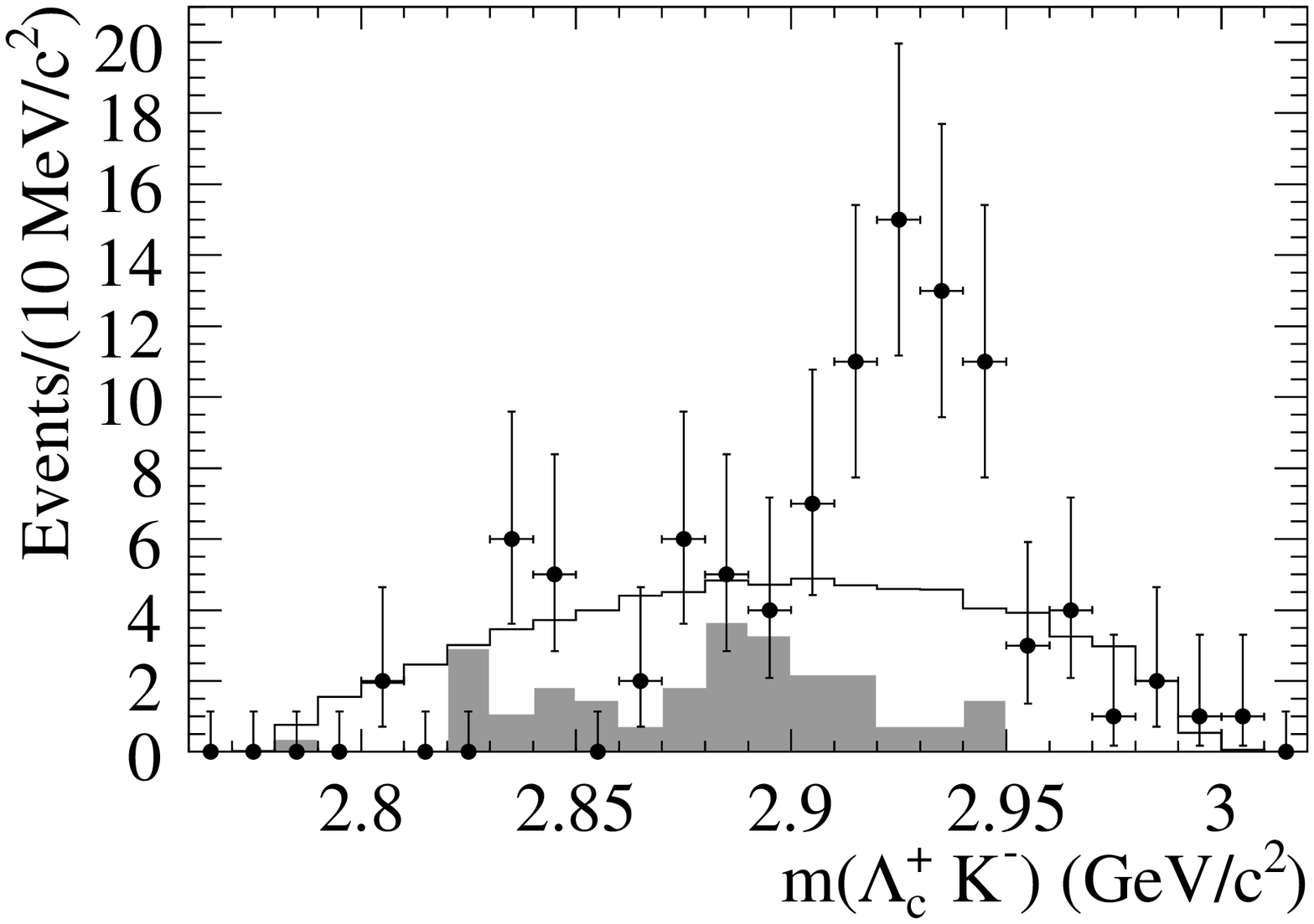, width=0.32\textwidth} 
      \begin{picture}(0,0)
	\put(-31, 95){(b)}
      \end{picture}
      &
      \epsfig{file=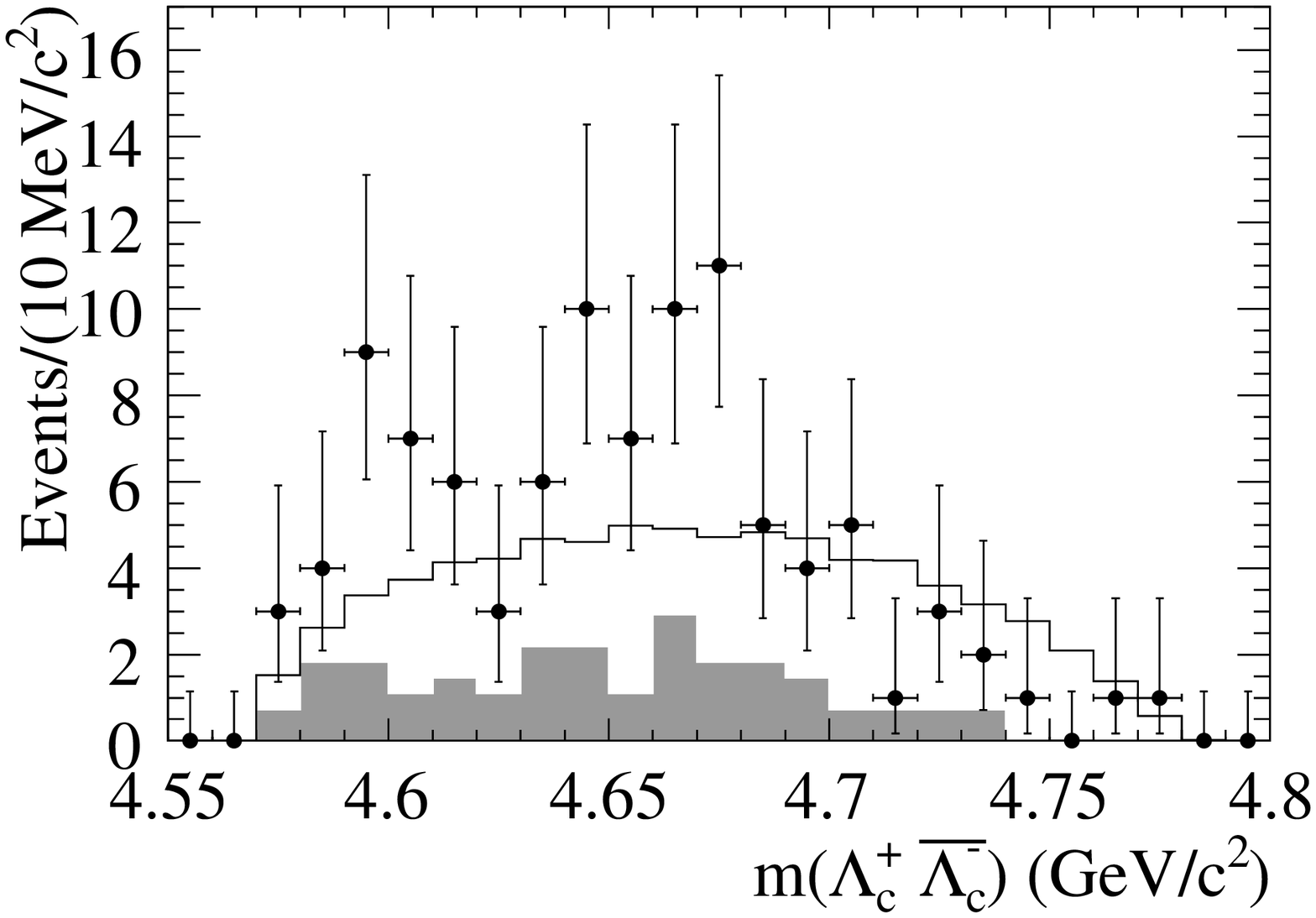, width=0.32\textwidth} 
      \begin{picture}(0,0)
	\put(-34, 95){(c)}
      \end{picture}
      \end{tabular}
  \end{center}
  \caption{
    Reconstructed \BmtoLcLcKm\ candidates in the signal region ($\mes >5.27\gev/c^2$, $\Delta E < 22$~MeV), shown as (a)~the Dalitz plot, (b)~the $\Lc\Km$ invariant mass distribution, and~(c) the $\Lc\aLc$ invariant mass distribution. Data from the signal region are shown as black points. Signal events from a phase-space simulation are shown as small grey points in~(a) and as a histogram in~(b) and~(c). Data from the sideband region $5.20 < \mes < 5.26$~GeV$/c^2$ are shown as a shaded histogram in~(b) and~(c), normalized according to the expected background yield in the signal region. The masses of the $B$-meson candidates are not constrained.
  }
\label{fig:dalitz}
\end{figure*}

Table~\ref{tab:systematic} lists the main systematic uncertainties and their sum in
quadrature. The largest uncertainty is from the charged track reconstruction efficiency, evaluated with control samples of $\tau$ decays. A small correction is also included due to a known data/MC difference in tracking efficiency. Other sources of systematic uncertainty considered include:
 the number of $\B\Bb$ pairs in the data sample;
 the limited size of the signal MC samples;
 the PID efficiency, which is evaluated with control samples of $\Lambda \to p \pi^-$, $\Dstarp \to D^0 (K^- \pi^+) \pi^+$, and $\phi \to K^+ K^-$ decays;
 possible differences in $\DeltaE$ resolution between data and MC, which are estimated with control samples of $\Bbar \to D \overline{D} \Kb$ decays;
 charmed baryon branching ratios relative to the control modes~\cite{ref:PDGbook};
 the \Lz branching fraction~\cite{ref:PDGbook};
   the presence of intermediate resonances in the charmed baryon decay
   and possible structure in the 3-body $B$-meson decays;
 and the assumption that $\BR(\Upsilon(4S)\to\BzBzb) = \BR(\Upsilon(4S)\to\BpBm) = 0.5$.
 For fit parameters which are fixed to values from fits to the signal MC, we vary the value by the uncertainty and take the largest change as a systematic uncertainty.
 Dividing out the absolute \Lc branching fraction also introduces a large systematic uncertainty, which we quote separately.

\begin{table}
\begin{center}
\caption{Summary of relative systematic uncertainties (\%) on the branching fractions. The uncertainty on the \Lc\ branching fraction is 26\% and is quoted separately.
}
\label{tab:systematic}
\begin{tabular} {lcccc} \hline \hline  \noalign{\vskip1pt}
Source & $\Lc\aLc K^-$ & $\Xi_c^0 \aLc$ & $\Xi_c^+ \aLc$ & $\Lc\aLc\Kzb$ \\
\hline
 Tracking efficiency          & ~9.9 & 10.0 & 11.4 & 11.4 \\ 
 \B counting                  & ~\,1.1 & ~\,1.1 & ~\,1.1 & ~\,1.1 \\
 MC sample size               & ~\,0.8 & ~\,1.6 & ~\,2.4 & ~\,1.5 \\
 PID efficiency               & ~\,4.6 & ~\,3.5 & ~\,3.0 & ~\,4.0 \\
 \DeltaE resolution           & ~\,3.0 & ~\,3.0 & ~\,3.0 & ~\,3.0 \\
 Intermediate BFs             & ~\,3.4 & ~\,6.9 & ~\,0.8 & ~\,0.1 \\
 $\Lc \to p K^- \pi^+$ Dalitz & ~\,2.9 & ~\,1.8 & ~\,1.8 & ~\,3.6 \\
 $B \to \Lc \aLc K$ Dalitz    & ~\,6.9 & ~\,--- & ~\,--- & ~\,4.2 \\
 $\Upsilon(4S)$ BF            & ~\,3.0 & ~\,3.0 & ~\,3.0 & ~\,3.0 \\
 Fit related                  & ~\,2.0 & ~\,1.4 & ~\,3.5 & ~\,2.5 \\
 \hline
 Total                        & 14.5 & 13.7 & 13.4 &14.3 \\
\hline \hline
\end{tabular}
\end{center}
\end{table}

To investigate whether the three-body mode \BmtoLcLcKm contains intermediate
resonances, we examine the Dalitz plot structure of candidates in the signal
region ($\mes >5.27\gev/c^2$), shown in Fig.~\ref{fig:dalitz}. After taking
into account the expected background (estimated from the \mes sidebands),
the $\Lc K^-$ mass spectrum of the data is inconsistent with a
phase-space distribution ($\chi^2$ probability of $1.5 \times 10^{-7}$). 
Fitting the data with a single, non-relativistic Breit-Wigner lineshape
convolved with a Gaussian function for experimental resolution, we obtain
  $m=2931 \pm 3 \mathrm{(stat)} \pm 5 \mathrm{(syst)}$~MeV$/c^{2}$ and
  $\Gamma=36 \pm 7 \mathrm{(stat)} \pm 11 \mathrm{(syst)}$~MeV.
We do not see any such structure in the \mes sideband region.
This description is in good agreement with the data ($\chi^2$ probability
of 22\%) and could be interpreted as a single $\Xi_c^{0}$ resonance
with those parameters, though a more complicated explanation (e.g. two
narrow resonances in close proximity) cannot be excluded. 
Due to the limited statistics, the helicity angle distribution does not
distinguish between spin hypotheses.

 In summary, we have studied $B$-meson decays to charmed baryon pairs in four decay modes using a sample of 230~million \upsbb events. 
The branching fraction of \BmtoLcLcKm is found to be larger than the previous measurement~\cite{ref:threebodycharm} and is comparable to the $\mathcal{O}(10^{-3})$ branching fraction predicted for two-body decays to a pair of charmed baryons. The other results are consistent with the previous values~\cite{ref:threebodycharm,ref:twobodycharm}. The data in the Dalitz plot and two-body mass projections of \BmtoLcLcKm are inconsistent with a phase-space distribution and suggest the presence of a $\Xi_c^{0}$ resonance in the decay. 

We are grateful for the 
extraordinary contributions of our \pep2\ colleagues in
achieving the excellent luminosity and machine conditions
that have made this work possible.
The success of this project also relies critically on the 
expertise and dedication of the computing organizations that 
support \babar.
The collaborating institutions wish to thank 
SLAC for its support and the kind hospitality extended to them. 
This work is supported by the
US Department of Energy
and National Science Foundation, the
Natural Sciences and Engineering Research Council (Canada),
the Commissariat \`a l'Energie Atomique and
Institut National de Physique Nucl\'eaire et de Physique des Particules
(France), the
Bundesministerium f\"ur Bildung und Forschung and
Deutsche Forschungsgemeinschaft
(Germany), the
Istituto Nazionale di Fisica Nucleare (Italy),
the Foundation for Fundamental Research on Matter (The Netherlands),
the Research Council of Norway, the
Ministry of Science and Technology of the Russian Federation, 
Ministerio de Educaci\'on y Ciencia (Spain), and the
Science and Technology Facilities Council (United Kingdom).
Individuals have received support from 
the Marie-Curie IEF program (European Union) and
the A. P. Sloan Foundation.


\begin{thebibliography}{19}
\bibitem{charge_conjugation}
Throughout this paper, for any given mode, the corresponding charge-conjugate mode is also implied.
\end{thebibliography}
\end{document}